\documentclass[preprint2,longabstract]{aastex}

\shorttitle{Envelope Around HH 211-mm}
\shortauthors{Tanner \& Arce}

\usepackage{parskip}

\begin{document}

\setcounter{figure}{0}

\title{The Dynamics of the Envelope Surrounding the Protostar HH 211-mm}
\author{Joel D. Tanner}
\affil{Department of Astronomy, Yale University, P.O. Box 208101, New Haven CT 06520}
\email{joel.tanner@yale.edu}

\and

\author{H\'{e}ctor G. Arce}
\affil{Department of Astronomy, Yale University, P.O. Box 208101, New Haven CT 06520}
\email{hector.arce@yale.edu}

\begin{abstract}

We present a study of the structure and dynamics of the dense gas surrounding the HH 211-mm source, using VLA observations of the ammonia (1,1) and (2,2) inversion transitions.  We find the envelope around this Class 0 source has an elongated geometry, extending about $10^4$ AU in the direction perpendicular to the well-known HH 211 jet, and exhibits a velocity distribution consistent with rotation along the major axis. Our VLA observations indicate that the envelope is mostly in virial equilibrium. However, comparing our data with results from previous studies, it appears that the gas within approximately 0.005 pc of the central protostar is undergoing dynamical collapse. The size of this collapsing radius may constrain the amount of mass that can eventually infall into the forming star. We also find that the envelope is mostly internally heated, most probably by radiation from the central protostar. In addition, we detect evidence of outflow-envelope interaction in the ammonia data. These include a velocity gradient in the dense gas along the outflow axis and significant line broadening that is spatially correlated with the jet and could be the result of outflow-induced turbulence in the envelope.

\end{abstract}


\section{Introduction}

Protostars form as a consequence of the gravitational contraction of dense gas in molecular cloud cores.  Once the process of star formation begins, the circumstellar dense gas in the core will act as the primary mass reservoir of the forming star.  Studying the physical and chemical properties of cores at different evolutionary stages is therefore important for understanding how protostar and protoplanetary disks form inside them.  In low-mass star-forming regions, cores are typically a few $10^4$ AU in size and can exhibit complex structures and dynamics.  Different velocity patterns (i.e., infall, outflow, rotation, and turbulence) may exist in the dense circumstellar gas. Inside cores, gravitational collapse will produce over-dense regions surrounding the forming star. However, rotation and turbulence can trigger fragmentation, thereby producing condensations where more protostars could form in the core.  In order to study the properties of dense cores in detail, observations with relatively high angular and velocity resolution are needed to disentangle the complex density and kinematic distribution that may be present in the dense circumstellar gas.

Here we present a study of the dense gas surrounding HH 211-mm, a young Class 0 source. 
This embedded protostar is the driving source of the well-known jet and highly-collimated molecular outflow HH 211 \citep{mccaughrean94, gueth1999}. Recent observations indicate that HH 211-mm may actually comprise two small sources (SMM1 and SMM2), with a total mass of about 0.05 M$_\odot$ (Lee et al. 2009).  This source lies in an area of dense gas southwest of the main IC 348 young stellar cluster, also known as the  IC 348-SW region (Bachiller et al 1987; Tafalla et al. 2006).  Distances to this region, which is part of the Perseus molecular cloud complex, range from 230 pc \citep{cer90} to 350 pc \citep{her83}.  Here we adopt a distance of 280 pc, similar to that used in one of the latest studies on this source (Lee et al. 2009). 

We use the line emission from ammonia to study physical properties of the dense gas surrounding HH 211-mm. This molecule is a high-density tracer which conveniently has more than one observable inversion transition line near a frequency of 23 GHz, that can be used to reliably estimate the temperature, kinematics and column density of the gas (e.g., Ho \& Townes 1983). This paper is the first of a series focused on understanding the evolution of the dense gas in cores using ammonia observations from the Very Large Array (VLA).

\section{Observations}

Observations were carried out with the VLA of the National Radio Astronomy Observatory\footnote{The National Radio Astronomy Observatory is a facility of the National Science Foundation operated under cooperative agreement by Associated Universities, Inc.} on December 13, 2005.  The NH$_3$ $(J,K) = (1,1)$ and $(2,2)$ inversion transitions, with  rest frequencies of 23.694495 and 23.722633 GHz, respectively, were observed simultaneously.  The array was in the compact (D) configuration. The bandwidth was $1.56$ MHz, and the channel separation was 12.2 kHz (corresponding to 0.154 km s$^{-1}$). We used the quasar 0319+415 with a flux density of $10.6$ Jy at $1$ cm as the absolute flux calibrator, and the quasar used for phase and amplitude calibration was 0336+323.  The raw data were reduced using NRAO AIPS image processing software.  The signal from each baseline was inspected, and baselines showing spurious data were removed prior to imaging.  The images were created using the IMAGR task with a robust parameter of $2$ and a cleaning beam width of $4\arcsec$.    Each channel was cleaned separately according to the spatial distribution of the emission.

\section{Results and Discussion}

\subsection{Spectral Modeling}
\label{sec:spectralmodeling}

\begin{figure}[h]
\epsscale{1.00}
\plotone{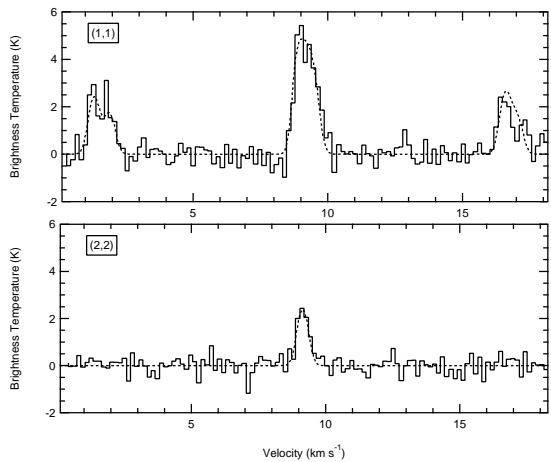}
\caption{Observed spectra of the ammonia (1,1) and (2,2) inversion transitions towards the position of HH 211-mm, averaged over one beam width 
($4\arcsec$).  The main and two satellite groups are visible in the (1,1) transition, while only the main group is visible for the (2,2) transition. 
Solid line denotes observed spectra, while the doted line shows the fitted model.
}
\label{fig:sampledata}
\end{figure}

The  $\mbox{NH}_3$ inversion spectrum is composed of a number of magnetic hyperfine components (Rydbeck et al.~1977; Ho \& Townes 1983).  In Figure \ref{fig:sampledata} we show an example (1,1) and (2,2) ammonia spectra for HH 211-mm.  The spectra are averaged over one synthesized beam width that is centered on the source.  In the (1,1) spectrum, we were able to detect the main quadrupole hyperfine and two satellite groups across our entire map (see the sample spectrum in Figure \ref{fig:sampledata}). In the case of the (2,2) spectrum, only the main hyperfine group is detected, although the satellite groups are within the passband.  

To extract physical parameters from the spectra, we model the (1,1) and (2,2) in a similar manner to \citet{rosolowsky2008}.  This modeling technique provides us with statistical fitting errors that we use as the uncertainty in the modeled quantities.  The brightness temperature ($T_B$) can be modeled with the radiation excitation temperature and the opacity distribution:
\begin{equation}
T_B = \left( J(T_{ex})  - J(T_{bg}) \right) \left( 1 - e^{-\tau(\nu)} \right),
\end{equation}
where $T_{\rm{ex}}$ is the radiation excitation temperature, $T_{\rm{bg}}=2.73$ K and 
\begin{equation}
J(T) = \frac{h \nu}{k} \frac{1}{\exp(h \nu / k T) - 1}.
\end{equation}

The opacity distribution ($\tau(\nu)$) for both spectra is determined by the total opacity ($\tau_1^{\rm{tot}}$ and $\tau_2^{\rm{tot}}$) and the opacity weights for each of the magnetic hyperfine components in the spectrum.  The opacity function is modeled with a superposition of Gaussians corresponding to each of the magnetic hyperfine components.  Previous studies have shown that representing the ammonia spectra in such a way provides very good fits to spectra with high and low opacity \citep{benson1989}. 
The contribution from each component is defined by its relative frequency ($\nu_i$), opacity weight, line width, and the line frequency associated with the LSR velocity ($\nu_{\rm{LSR}}$).  All hyperfine components are constrained to have the same line width.  The opacity distribution for the (1,1) and (2,2) spectra are modeled separately with:
\begin{equation}
\tau(\nu) = \tau^{tot} \sum_i^N \left( s_i \exp(- \frac{\nu - \nu_i - \nu_{\rm{LSR}}}{2 \sigma}), \right)
\end{equation}
where $s_i$ is the statistical weight of a particular magnetic hyperfine component, $\tau^{tot}$ is the total opacity ($\tau_1^{tot}$ or $\tau_2^{tot}$ for the (1,1) or (2,2) spectra, respectively) and $N$ is the number of magnetic hyperfine components in the spectrum.  We use the statistical weights from the CLASS\footnote{http://www.iram.fr/IRAMFR/GILDAS} software package for both the (1,1) and (2,2) models.

Using a non-linear least-squares minimization \citep{marqwardt2009}, we find the best fit to the observed spectrum at all points.  The (1,1) spectrum is modeled with four free parameters: the total opacity ($\tau_1^{tot}$), excitation temperature ($T_{ex}$), line width ($\Delta v$) and LSR velocity ($v_{\rm{LSR}}$).  We determine most of the parameters from the (1,1) spectral model because the observed (1,1) data has a substantially higher signal-to-noise ratio than the (2,2). After determining the parameters of the (1,1) spectral model, the (2,2) spectrum is modeled in a similar manner, but using $T_{ex}$, $\Delta v$, and $v_{\rm{LSR}}$ from the (1,1) model, and so is determined with the single parameter $\tau_2^{tot}$.  

Once the fitting parameters have been extracted from the model spectra, the rotational excitation temperature is easily computed with the ratio of the (2,2) to the (1,1) optical depths using:
\begin{equation}
T_{rot} \approx -41 \mbox{K} / ln \left( \frac{9}{20} \frac{\tau_{2}^{tot}}{\tau_{1}^{tot}} \right) .
\end{equation}
This formulation for rotation temperature requires modeling both the (1,1) and (2,2) spectra.  However, the area where we detect (2,2) emission is smaller than that with (1,1) emission (see \S~\ref{sec:morphology}).  To estimate the column density in regions with only (1,1) emission we use the value of $T_{rot}$ from the regions where (2,2) emission is at the threshold of detection (i.e., $T_{rot} \sim 8.6$~K).

Column density is obtained by integrating the opacity over velocity space.  From \citet{mangum1992}, the column density of the (1,1) and (2,2) transitions is:
\begin{equation}
N(1,1) = 6.60 \cdot 10^{14} \frac{T_{rot}}{\nu(1,1)} \tau(1,1,m)\Delta v \mbox{\hspace*{0.3cm}cm}^{-2};
\end{equation}
\begin{equation}
N(2,2) = 3.11 \cdot 10^{14} \frac{T_{rot}}{\nu(2,2)} \tau(2,2,m)\Delta v \mbox{\hspace*{0.3cm}cm}^{-2},
\end{equation}
where $\Delta v$ is the line width in $\rm{km} \rm{s}^{-1}$, and $\nu$ is the frequency of the hyperfine transition in GHz.  The total column density is then calculated with
\begin{equation}
N_{tot} = \frac{N(J,K) Q_{rot}}{(2J+1)g_I g_K} exp \left[ \frac{E(J,K)}{T_{rot}}  \right],
\end{equation}
where $g_I$ ($= 2$ for $K \ne 0$ or $3n$) is the nuclear spin degeneracy, $g_K$ ($= 2$ for $K \ne 0$) is the K-degeneracy,  and $E(J,K)$ is the energy of the inversion state above the ground state.  The rotational partition function is given by \citep{gordy1970}:
\begin{equation}
Q_{rot} \approx \frac{5.34\cdot 10^6}{3} \left[ \frac{T_{rot}^3}{CB^2} \right]^{1/2}, 
\end{equation}
where the rotational constants B and C are $298117$ and $186726$ $\rm{MHz}$, respectively.

The temperature of the envelope is considerably less than $T_o = 41.5$ K, reaching temperatures in excess of $20$ K only in regions very near the central source.  In this low temperature regime we can solve the equation from \citet{swift2005} to obtain the kinetic temperature.
\begin{equation}
T_{rot} = T_k \left[ 1 + \frac{T_k}{T_{0}} \ln \left( 1 + 0.6 e^{-15.7/T_k}  \right) \right] ^{-1}
\end{equation}

\subsection{Envelope Morphology and Mass}\label{sec:morphology}

\begin{figure}[h]
\epsscale{1.00}
\plotone{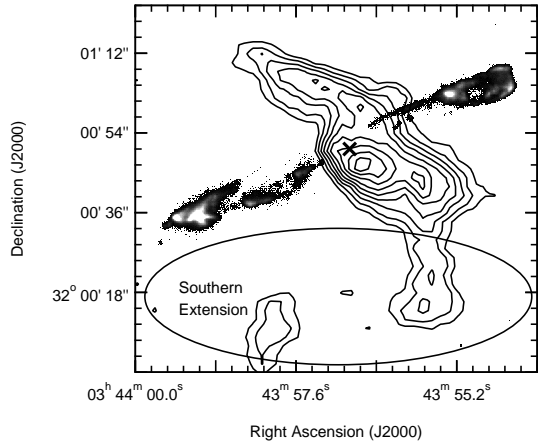}
\caption{Integrated intensity map of the ammonia (1,1) transition with signal-to-noise contours starting at $3 \sigma$ in increments of $2 \sigma$ ($1.95$ mJy/Beam).  The $\mbox{NH}_3$ data are represented with contours and is superimposed on an $\mbox{H}_2$ image of the jet (from Hirano et al. 2006).  The location of the source is marked with a cross. The extent of the southern extension of the envelope is shown.
} 
\label{fig:intensity_all}
\end{figure}

An integrated intensity  (0th moment) map of the central (1,1) emission is shown in Figure \ref{fig:intensity_all}.  The intensity contours are shown at various detection strengths above the noise, separated by $2 \sigma$ and are superimposed on the $\mbox{H}_2$ $\nu=1-0$ $S(1)$ line at 2.122 {\micron} (narrow-band filter) map of \citet{hirano2006}, which shows the HH 211 protostellar jet.  The NH$_3$ (1,1) emission traces an elongated structure that is perpendicular to the jet outflow axis and shows a peak in integrated intensity approximately 2.5" ($\sim 5/8$ of a synthesized beam width) from the protostellar source HH 211-mm, consistent with the NH$_3$ (1,1) integrated intensity map of the same region shown by Wiseman (2001). 

We observe ammonia emission tracing an extension of gas eastward from the southern end of the envelope with low signal-to-noise ratio. Its average LSR velocity is about 0.5 km~s$^{-1}$ blueshifted from the average LSR velocity of the southern part of the main envelope, and it has a very different structure and steeper velocity gradient than that of the rest of the envelope. This southern extension accounts for approximately 10\% of the envelope mass (assuming a rotation temperature of $10$ K), and from its structure and velocity distribution it appears to be only very loosely related to the rest of the core. We therefore neglect it throughout most of our subsequent analysis.  

The geometry of the envelope is believed to be an elongated structure seen nearly edge-on.  In their recent study, \citet{lee2009} determined that the HH 211 molecular outflow has an inclination to the plane of the sky ($\alpha$) of about $5\arcdeg$, using the measured radial and traverse velocities of the collimated molecular outflow. Assuming that the mid-plane of the elongated ammonia envelope is perpendicular to the jet axis ---as expected from results of different collapse models (e.g., Galli \& Shu 1993; Hartmann et al.~1996)--- then the observed ammonia structure should be approximately edge-on. 

The extent of the detected $\mbox{NH}_3$ (1,1) emission spans approximately $1.8 \times 10^4$  AU along the major axis  (assuming a distance of 280 pc to the source).  To extract a characteristic radius for this structure  we fit a 2D Gaussian to the integrated $\mbox{NH}_3$ (1,1) intensity, similar to the procedure used by previous studies for obtaining sizes of ammonia or N$_{2}$H$^+$ cores   \citep[see, e.g.,][]{benson1989,chen2007}. The fit defines a major and minor axis, which are nearly perfectly perpendicular and parallel to the jet outflow axis, respectively.  We measure the elongated core axis ratio of approximately 3:1, with the half width of the major axis to be approximately $6200$~AU. From this fit, we estimate the thickness of the  nearly edge-on structure to be approximately 4000 AU. This structure is too large and ``puffy'' to be a circumstellar disk, and most likely traces the extended dense envelope surrounding the HH 211-mm protostar. As we discuss below, there is a clear velocity gradient along the major axis of the core, possibly due to rotation of the extended envelope around HH 211-mm. The elongated morphology and velocity structure observed in the dense gas within $\sim 10^4$ AU of HH 211-mm appears to be a relatively common characteristic (although not necessarily universal, see, Tobin et al. 2010) of the dense gas surrounding low-mass young and embedded protostars, as a number of Class 0 sources that have been observed at high angular resolution show similar structures in ammonia lines and other molecular line high-density tracers \citep[e.g.,][]{wiseman2001,belloche2002,chen2007,chiang10}.

\begin{figure}[h]
\epsscale{1.00}
\plotone{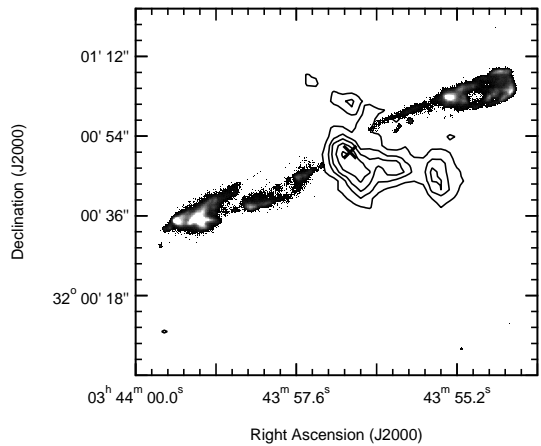}
\caption{Similar to Fig. \ref{fig:intensity_all} but showing the integrated intensity of the ammonia (2,2) transition with signal-to-noise contours starting at $3 \sigma$ and in increments of $1 \sigma$ (1.75 mJy/Beam).} 
\label{fig:intensity22}
\end{figure}

Figure \ref{fig:intensity22} shows the integrated intensity map for the  $\mbox{NH}_3$ (2,2) transition.  Although the detected (2,2) emission does not extend over the same region as the (1,1) emission, the (2,2) emission follows the distribution of the high signal-to-noise (1,1) emission and peaks near source. There is a low signal-to-noise detection to the north of the source which we include in the analysis of the envelope, but note that the temperature and column density estimates for this part of the map are less certain than those near the source.

\begin{figure}[h]
\epsscale{1.00}
\plotone{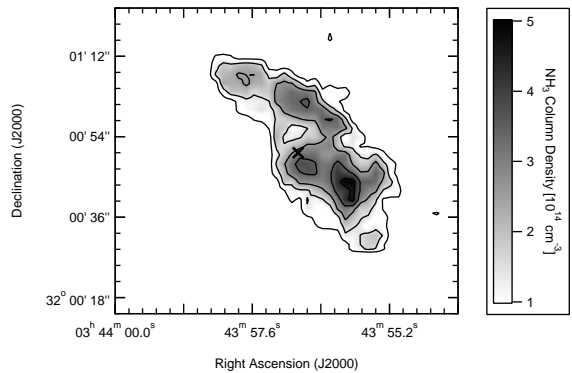}
\caption{Ammonia column density map. Contours are in steps of  $1.0 \times 10^{14}$ and start at $0.5 \times 10^{14}$  $\rm{cm}^{-2}$.  There is a clear local peak near the central source that extends to the south west, and two other condensations to the north and south. } 
\label{fig:coldens}
\end{figure}

In Figure \ref{fig:coldens} we show a map of the $\mbox{NH}_3$ column density, obtained using  the procedure described in Section \ref{sec:spectralmodeling} for each  2" pixel over the entire map.
The resulting $1 \sigma$ statistical errors from the spectral fitting are nearly uniform at $\sim$ $10$\% across the map. The measured column density range from less than $10^{14}$ cm$^{-2}$ at the edges of the core to approximately $5 \times 10^{14}$ cm$^{-2}$. This is consistent with the value ($4 \times 10^{14}$ cm$^{-2}$) estimated by \citet{rosolowsky2008} using single-dish  ammonia line observations (with a $30\arcsec$ beam) toward the center of the core (source NH3SRC 161 in their Table 3). 
 
The column density map in Figure~\ref{fig:coldens} shows three local column density peaks with values larger than $3.5 \times 10^{14}$ cm$^{-2}$  distributed over the area of the NH$_3$ envelope.  The local peak approximately $2.5\arcsec$ from HH 211-mm is probably associated with the dense inner envelope where the Class 0 source formed. It is possible that the other two peaks in column density are the result of  fragmentation within the envelope, which could have been caused by the relative fast rotation in the core (see \S~\ref{sec:rotation}). Using an ammonia to hydrogen abundance ratio of $3 \times 10^{-8}$ \citep{bachiller1987} and a mean molecular weight of $2.74$ (which accounts for 10\% helium), the total mass of the observed ammonia envelope is 0.4 M$_\odot$, which is approximately a factor of  $8$ larger than the mass of the central protostar as estimated by \citet{lee2009}.

\subsection{Rotation Signature}\label{sec:rotation}

\begin{figure}[h]
\epsscale{1.00}
\plotone{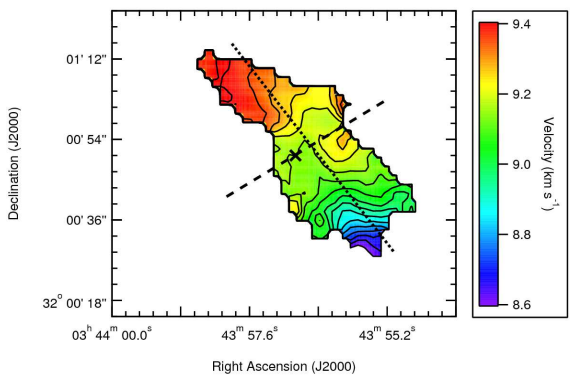}
\caption{First moment map of the $\mbox{NH}_3$ (1,1) emission.   The dotted and dashed lines show the major and minor axes of the envelope, along which we measure the velocity gradients (see Figures \ref{fig:meanvelgrad} and \ref{fig:velgrad_minor}).  The position of HH 211-mm is marked with a cross. Contours start at LSR velocity $8.6~\rm{km}~\rm{s}^{-1}$ in increment in steps of $0.05$ $\rm{km} ~\rm{s}^{-1}$.} 
\label{fig:velmap}
\end{figure}

Figure \ref{fig:velmap} shows the velocity field (1st moment map) of the ammonia emission.  As in Figure \ref{fig:intensity_all}, the contours are superimposed on the $\mbox{H}_2$ jet image. We clearly see that the radial (LSR) velocity changes smoothly from redshifted in the north-east end of the envelope, to blueshifted in the south-west. The direction of the mean velocity gradient is almost perfectly aligned with the long axis of the core, and is approximately perpendicular to the jet outflow axis.  

\begin{figure}[h]
\epsscale{1.00}
\plotone{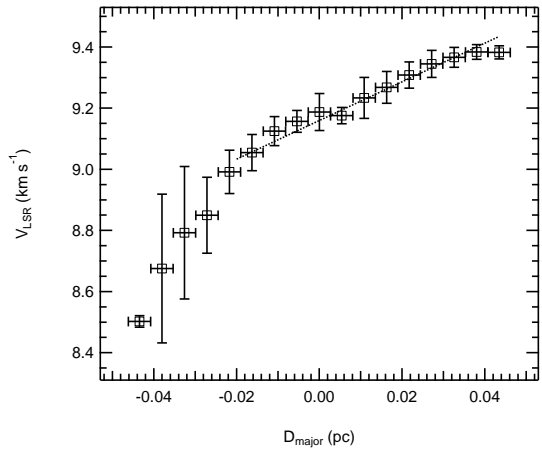}
\caption{Velocity as function of  radius measured along the major axis (see Figure \ref{fig:velmap}) from south (negative values) to north (positive values) with the zero point aligned with the position of HH 211-mm.  The velocity shown is the (LSR) velocity component along the line-of-sight.
 The plotted values and error bars represent the mean and standard deviation of measured velocities within bins that are $2\arcsec$  (half a beam width) wide and cover the extent of the envelope perpendicular to the major axis.  The mean velocity gradient for the envelope (dashed line) is calculated excluding velocities affected by the southern extension (i.e., the first four points on the left of the plot).} 
\label{fig:meanvelgrad}
\end{figure}

The mean radial velocity along the major axis (averaged over the minor axis of the envelope) is shown in Figure \ref{fig:meanvelgrad}, where the zero point is set at the position of the central source along the major axis, and negative and positive offsets come from positions southeast and northwest of the source, respectively. The ammonia emission exhibits a smooth and nearly constant gradient along most of the major axis, with distinct deviations that are discussed in more detail in Section \ref{sec:outflow}. A linear fit to most of the points in Figure \ref{fig:meanvelgrad} gives a gradient ($g$) of approximately $6.3$ km s$^{-1}$ pc$^{-1}$. We did not use the southernmost four points, as they are affected by the kinematics of the southern extension (see \S~\ref{sec:morphology}).

The nearly constant velocity gradient along the major axis of the envelope suggest that the dense gas within about 6000 AU of the protostar  has a  rigid rotation velocity structure, rather than a Keplerian dominated profile.  The non-Keplerian profile of the extended envelope, and the relatively large envelope mass of about an order of magnitude greater than the total mass of the central sources (from Lee et al. 2009), are consistent with the characteristics of a young protostellar (Class 0) system in which most of the mass still resides in the envelope.  

Fast rotation in the envelope may affect the distribution of the dense gas by causing fragmentation or flattening of the rotating structure.  We quantify the amount of rotation  with the parameter $\beta_{rot}$, which is defined as the ratio of the rotational kinetic energy ($E_{rot}$) to the gravitational energy ($E_{grav}$).  The geometry and density stratification of the circumstellar gas determine both the moment of inertia and the gravitational potential.  We approximate the structure of the HH 211 envelope to be a uniform density cylinder with $E_{rot} = (1/2) I \omega^2$ and $E_{grav} = (2/3) GM^2/R$,  where $I = (1/2)M R^2$ is the moment of inertia, $\omega$ is the angular rotation speed, $G$ is the gravitational constant and $M$ and $R$ are the mass and radius of the cylinder. The value of  $\beta_{\rm rot}$ for the HH 211 envelope is given by:
\begin{equation}
\beta_{\rm rot} = \frac{E_{rot}}{E_{grav}} = \frac{3}{8} \frac{ \omega^2 R^3}{GM}.
\end{equation}
We  assume that the envelope is perpendicular to the jet (and has an inclination of $i \sim 85\arcdeg$) so that $\omega = g / \sin i \approx g \approx 6.3 \mbox{ km} \mbox{s}^{-1} \mbox{pc}^{-1}$, and adopt the envelope radius to be the semi-major axis from the Gaussian fit to the integrated intensity (see \S~\ref{sec:morphology}). Using these values, we estimate the extended NH$_3$ envelope surrounding HH 211-mm to have a value of $\beta_{rot}$ of approximately 0.2. This is larger than the values observed in the sample of protostellar envelopes studied by \citet{chen2007}, but comparable to the values observed in the envelope of the Class 0 source IRAM 04191 \citep{belloche2002}.   Small values of $\beta_{rot}$ indicate that the rotational kinetic energy of the envelope is small compared to its gravitational binding energy, while systems in which rotation plays a significant role in maintaining  envelope equilibrium against collapse will have large values of $\beta_{rot}$ \citep{tohline2002}.  

The rotational energy in the dense gas surrounding HH 211-mm is relatively high, and could be the cause of the envelope's observed flattened morphology. Numerical studies have shown that rotating, magnetic clouds will experience fragmentation when $\beta_{rot}$  is at least 0.01 to 0.05 (Boss 1999; Machida et al. 2005). The high $\beta_{rot}$ we observe in our source may have caused the presumed fragmentation in the envelope that produced the northern and southern peaks in the ammonia column density map (seen in Figure \ref{fig:coldens}).

\subsection{Gas Temperature}

\begin{figure}[h]
\epsscale{1.00}
\plotone{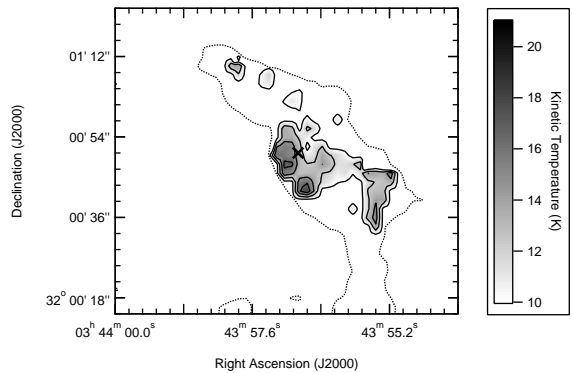}
\caption{Kinetic temperature map of gas surrounding HH 211-mm.  The region near the central source shows heating in excess of 20 K and coincides with the column density peak that extends from the source to the south-west (see Figure \ref{fig:coldens}).   The dotted line shows the extent of the detected (1,1) emission.  The kinetic temperature is estimated with an assumed rotational temperature of 8.6 K in regions where only (1,1) emission is detected. Contours are in steps of $3$ K and start at $10$ K.} 
\label{fig:tkin}
\end{figure}

The ratio of the NH$_3$ (1,1) to (2,2) emission is determined by the level populations, and can be used to estimate the rotational and kinetic temperature of the gas within the molecular core (see \S \ref{sec:spectralmodeling}).  Figure \ref{fig:tkin} shows a grey scale map of the kinetic temperature of the gas superimposed on the outline of the envelope traced by the ammonia (1,1) integrated intensity map (Figure \ref{fig:intensity_all}). Our observations were not sensitive enough to detect (2,2) emission throughout the entire envelope, and  the map in Figure \ref{fig:tkin} only shows values of the temperature for regions where (2,2) was detected $3 \sigma$ above the noise. The $1 \sigma$ statistical errors of the temperature estimate are nearly uniform across the map, and are typically around $15 \%$. The temperature of the envelope ranges from 10 to 21 K, with a mean value of 15 K, consistent with the value obtained by \citet{rosolowsky2008} for NH3SRC 161.

The gas is not heated equally throughout, and shows a statistically significant local maximum (of about 20 K) close to the protostellar source.  The temperature profile  approximately dropps off with distance from the central source (see Figure \ref{fig:tkin}). Apart from regions close to the source, most of the gas with a temperature of 15 K or more resides away from the edges of the ammonia core, unlike externally heated pre-stellar cores (e.g., Friesen et al.~2009). The observed temperature distribution in the HH 211 envelope  suggests that the core is being internally heated. We do not detect any clear temperature increase along the outflow axis, northwest of the source. It is, therefore, most likely that the increase in temperature near the source seen in Figure \ref{fig:tkin} is caused by the radiation from the embedded protostar, which has a bolometric luminosity of 3.6~L$_{\sun}$ (Froebrich 2005).

Although the envelope is generally consistent with internal heating, the localized peaks in the temperature map that are more than $10\arcsec$ from the source could be the result of external heating.  It is plausible that shocks in the HH211 H2 are externally heating parts of the envelope.  In particular, the nearby knot `f' (from McCaughrean \textit{et al}. 1994), which lies to the north of the envelope (about $15\arcsec$ north-west of the source), has a temperature in the range of 2500-2800 K (Caratti o Garatti \textit{et al}., 2006) and could be responsible for the local peak in gas temperature that is about $ 20\arcsec$ south-west of the source.

\subsection{Line Widths}
 \label{sec:fwhm} 

\begin{figure}[h]
\epsscale{1.00}
\plotone{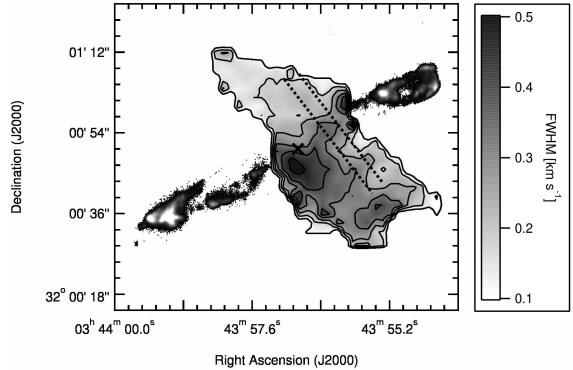}
\caption{Map of the observed NH$_3$ (1,1) line width.  Statistical errors are typically $12$\% of the measured value across the map.  Contours start at 0.1 $\rm{km}~\rm{s}^{-1}$ and are in increments of 0.07 $\rm{km}~\rm{s}^{-1}$. Dashed lines indicate the width of the slice used to obtain the velocity width profile, perpendicular to the outflow axis, shown in Figure~\ref{fig:fwhmslice}.} 
\label{fig:fwhm}
\end{figure}

For each position in the map with significant (1,1) emission, we measured the observed line width. 
Figure \ref{fig:fwhm} presents a map of $\Delta v$, the full width at half maximum (FWHM) of the ammonia emission line, from the spectral model fit to each pixel across the (1,1) map.  The $1 \sigma$ statistical errors in the observed line widths show no systematic variation across the map, and are typically $12$\% of the measurement.    The measured line widths range from about  0.15 to 0.48 km s$^{-1}$ over the entire core, and are the largest near the source, where the column density and rotational temperature are high.  We also detect relatively high line widths to the northwest of the source, coincident with the HH 211 jet, and  southwest of the source along the envelope's major axis. 

The observed total line width is due to a combination of thermal and non-thermal motions:
\begin{equation}
\Delta v^2_{obs} = \Delta v^2_{NT} + \Delta v^2_{therm},
\end{equation}
where the thermal line width is given by:
\begin{equation}
\Delta v^2_{therm} = 8 \ln 2 \frac{k T_K}{m_{obs}},
\end{equation}
where $k$ is the Boltzmann constant, $m_{obs}$ is the mass of the observed molecule ($17$ amu for $\mbox{NH}_3$) and $T_K$ is the kinetic temperature of the gas.  Similar to the observed line widths, the thermal width error estimates have no discernable variation but have relative errors typically around $8$\%.  The non-thermal motions are due to turbulence or unresolved bulk motions (such as rotation). 
Using the equations above, we see that in the core  $\Delta v_{NT}$ ranges from 0.03 to 0.46 km s$^{-1}$, and  $\Delta v_{therm}$ ranges from about 0.15 to 0.24 km s$^{-1}$. There are regions in the core, particularly in areas north-northeast of the central source, where the observed width is comparable to the thermal width and the non-thermal width is negligible. The line width in the envelope peaks close to the source, at a value of 0.48 km s$^{-1}$.  In this region, thermal broadening is not sufficient to account for the observed line widths, and a combination of higher gas temperatures and large non-thermal motions (possibly due to the outflow and an inner infalling region, see below) may contribute to the observed large velocity widths. The temperature map in Figure \ref{fig:tkin} shows that the envelope temperature is highest close to, and southwest of the source. Thus, the relatively high line width seen south-southwest of the source (with an average value of about 0.34 km s$^{-1}$) is partly due to the existing higher gas temperatures.  At the southwestern edge of the map, close to where the southern extension connects with the rest of the core, there is a local peak in the line width with a value of 0.38 km s$^{-1}$. As discussed above, the southern extension has a distinct velocity compared to the rest of the envelope.  The unresolved velocity component due to the southern extension emission results in apparent wider spectral lines at the southwestern edge of the envelope.  North-northwest, along the outflow axis, there is a narrow region with a relatively large velocity width of about 0.34 km s$^{-1}$. In this region the temperature is relatively low (see Fig.~\ref{fig:tkin}), and most of the line broadening is due to non-thermal motions. We speculate that this is due mostly to outflow-driven turbulence (see \S~\ref{sec:outflow}).

\subsubsection{Inner rotating disk?}

The high angular resolution observations of Lee et al. (2009) detect a flattened structure within $\sim 0.5''$ of the central source, with a velocity gradient consistent with a Keplerian or $v_{rot} \propto r^{-1}$ rotation.  Given the lengths of their projected baselines, their observations are insensitive to structures larger than about $4\arcsec$,   so that if the rotating structure were more extended than what is shown by Lee et al.~(2009), they would not have detected it. Observations taken with shorter baselines (from Lee et al.~2007) show that the continuum emission surrounding HH 211-mm along the major axis (perpendicular to the outflow axis) extends approximately  $4\arcsec$. Thus, it could very well be that the inner rotating structure surrounding the embedded protostar(s) has a radius of 2 to $3\arcsec$, or about  600 to 800 AU, which is not unusual for low-mass embedded protostars (e.g., Ohashi et al. 1997; Hogerheijde et al. 1998). If present,  such rotating structure would not be resolved in our synthesized map, but it would contribute to the observed line widths near the source.  

To investigate whether the possibility of an unresolved Keplerian disk near the central source is consistent with our data, we use a simple model to examine the effect that such a disk would have on the observations.  Our toy model disk is assumed to be edge-on and geometrically thin.  For a resolved geometrically thin disk with a Keplerian velocity profile, the  velocity across a disk defined by the polar coordinates $r$ and $\theta$ is \citep{guilloteau2006}:
\begin{equation}
v_{disk}(r,\theta) =  \left( \frac{GM}{r}  \right)^{1/2} \sin i \cos \theta ,
\end{equation}
where $i$ is the inclination angle of the disk (where edge-on disks have $i=90\arcdeg$), and the observed velocity is in a frame that is stationary with respect to the disk.  Contours of radial velocity are defined by:
\begin{equation}\label{eqn:contour}
r(\theta) = \frac{GM}{v_{disk}^2} \sin^2 i \cos^2 \theta .
\end{equation}
The line broadening will be determined by the length of the contours at a given velocity.  From equation \ref{eqn:contour} we see that the resulting profile is symmetric and will have a peak corresponding to the longest contour which will just touch the outer edge of the disk at $R=R_{out}$ and $\cos \theta = 0$.  We can define a characteristic broadening (see Guilloteau et al.~2006) by the longest isovelocity contour described by Eq.~14 as:
\begin{equation}
\Delta v_{Kepler} = 2 \sqrt{\frac{GM}{R_{out}}} \sin i .
\end{equation} 

According to \citet{lee2009}, if the gas motions near the source are  Keplerian, they then imply that the central object (or objects) have a total  mass of about $0.05 \pm 0.015$ M$_\odot$.  If we assume $i \sim 85\arcdeg$, a central mass of M$ = 0.05$ M$_{\odot}$, an outer radius consistent with half the synthesized beam of our map ($R_{out} = 2\arcsec$), and that there is NH$_3$(1,1) emission from the disk,  we then estimate that an unresolved Keplerian disk could contribute as much as $\Delta v_{Kepler} \sim 0.56 \mbox{ km s}^{-1}$ to the non-thermal broadening in the ammonia line emission
near the source position. This is significantly larger than the maximum line width observe near the source (0.48~km~s$^{-1}$). It appears that our ammonia observations do not probe the inner rotating structure close to HH 211-mm, either because the disk does not extend far beyond what is observed by  Lee et al.~(1999) (i.e., it is much smaller than our beam) or the ammonia line emission is not a good tracer of the disk, or both.    

Although the increase in line width close to the source is partially due to an increase in the thermal width, substantial non-thermal motions are needed to explain the observed line broadening. The thermal line width in this region is about 0.2~km~s$^{-1}$, and as we discuss below, the outflow appears to be responsible for a non-thermal broadening of the line ( $\Delta v_{outflow}$) of about 0.3~km~s$^{-1}$ along its axis. Adding these two in quadrature results in a total width of 0.36~km~s$^{-1}$, significantly smaller than the observed line width near the source. A possible source of additional non-thermal motions may come from infall motions close to the protostar. In  \S~\ref{sec:velocity} we suggest that the region within about 0.005 pc may be collapsing,  and unresolved infall motions may result in an increase line width. Another possibility is that $\Delta v_{outflow}$ increases near the source. Evidence for this is seen in the increase in the range of outflow velocities observed close to the source (Lee et al.~2007; 2009). Higher angular resolution observations that can probe the high density gas within $5\arcsec$ of the source are needed to further investigate the dynamics of the inner envelope.

\subsection{Outflow motions and turbulence} 
 \label{sec:outflow} 
\begin{figure}[h]
\epsscale{1.00}
\plotone{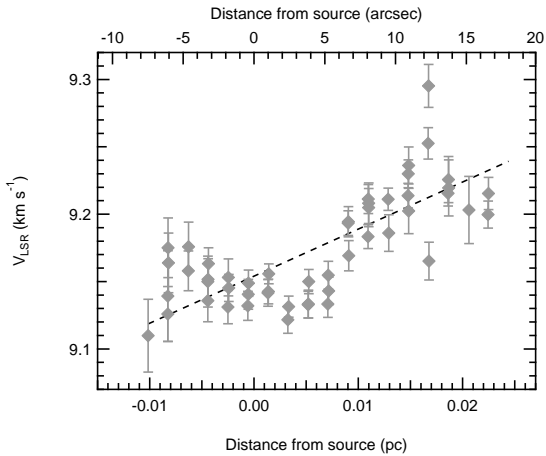}
\caption{The velocity gradient along the minor axis of the NH$_3$ envelope.  The gradient is measured over a $8\arcsec$ wide slice centered on the jet axis (see Figure \ref{fig:velmap}).  Each data point corresponds to a pixel from the velocity map that falls within the slice.  The error bars show the $1 \sigma$ statistical errors from the spectral fitting.  The dashed line is a linear fit to the data and has a gradient of approximately $3.5$ $\rm{km}~\rm{s}^{-1}~\rm{pc}^{-1}.$}
\label{fig:velgrad_minor}
\end{figure}

As discussed in Section \ref{sec:rotation}, and seen in Figure \ref{fig:velmap}, the velocity gradient of the envelope is mostly along the major axis of the envelope. However, there is a deviation from this gradient in the region close to the protostar and outflow axis, as the velocity contours here follow a distinct pattern compared to the contours in the envelope outskirts (see Figure \ref{fig:velmap}).
To further study this, we examine the velocity gradient along the minor axis of the envelope, which we present in Figure \ref{fig:velgrad_minor}. Each point in the figure represents the central (LSR) velocity from the spectrum in a $2\arcsec$ pixel, and the error bars indicate the $1\sigma$ in the velocity estimate (using the procedure described in \S~\ref{sec:spectralmodeling}). Only pixels within a strip that is $8 \arcsec$ wide (two beam widths) and centered on the jet axis are included. The plot shows that, along the outflow axis, most pixels northwest of the source are redshifted with respect to the central (source position) velocity, while most pixels southeast of the source are blueshifted.  A fit to the velocity gradient along the envelope minor axis yields a gradient of approximately $3.5$ km s$^{-1}$ pc$^{-1}$, increasing (from blue to red velocities) from  southeast to northwest of the source. 

We attribute the velocity distribution along the minor axis of the ammonia envelope to the effect of the HH 211 outflow on the dense gas surrounding the protostar. As mentioned above we are most likely viewing the ammonia envelope edge-on, so the gas southeast and northwest of the source (along the minor axis) correspond to regions above and below the mid-plane of the envelope. The outflow originates very close to the protostar (within $\sim 100$ AU, Lee et al.~2009), and must interact with the dense gas as it punctures through the envelope. The southeast lobe of the  HH 211 molecular outflow shows mostly blueshfited velocities, while the northwest lobe shows mostly redshifted velocities (Lee et al.~2007). Thus, it is very likely that outflow is pushing the dense gas as it moves away from the source---passing above and below the midplane of the envelope---and has produced the velocity gradient seen along the minor axis.  Similar velocity shifts in the dense envelope gas produced by outflows have been seen in a number of other Class 0 sources (e.g., Wiseman et al.~2001; Arce \& Sargent 2005, 2006).

If our interpretation is correct, then we can use the ammonia emission near the outflow axis to estimate the effects of the HH 211 jet on the dense gas. The mass of the ammonia envelope within the $8\arcsec$ wide slice used to obtain the points in Figure \ref{fig:velgrad_minor} is about 0.05 M$_{\sun}$ (excluding one beam width centered on HH 211-mm). From the velocity gradient along the minor axis shown in Figure \ref{fig:velgrad_minor} we estimate a ``characteristic'' velocity offset from the velocity at the source position ($\sim 9.15$ km~s$^{-1}$) of about 0.05 km~s$^{-1}$ along the outflow axis, measured half way between the source and the northwest edge of the envelope.  This provides a rough estimate of the momentum (0.0025 M$_{\sun}$ km~s$^{-1}$) and kinetic energy ($1.3 \times 10^{39}$ ergs) associated with the dense gas traced by the ammonia emission that has been pushed by the outflow (without correcting for outflow inclination).  These values are less than a factor of ten compared to the HH 211 molecular outflow momentum and kinetic energy (without inclination correction and assuming a distance of 280 pc) reported by Palau et al. (2006). In principle, then, the velocity gradient in the dense circumstellar gas along the minor axis could have been caused by the outflow. If we correct for the inclination of the outflow to the plane of the sky ($\alpha \sim 5\arcdeg$), our estimate of the average velocity of the  molecular outflow traced  by the NH$_3$ emission increases to about 0.6 km s$^{-1}$ and the  momentum and kinetic energy rises to   0.03 M$_{\sun}$ km s$^{-1}$ and $1.6 \times 10^{41}$ ergs, respectively. This velocity is higher than the escape velocity of the envelope ($\sqrt{2GM/R} \sim 0.3$~km~s$^{-1}$)  and the kinetic energy of the dense outflowing gas traced by the ammonia emission is comparable to  the gravitational binding energy of the envelope ($\sim 3 \times 10^{41}$ ergs).  On the basis of these estimates, we speculate that the protostellar wind has deposited enough energy to clear the ambient dense gas along the outflow axis and will eventually  produce outflow cavities, as seen in more evolved (Class I) embedded protostars (e.g., Arce \& Sargent 2006).

\begin{figure}[h]
\epsscale{1.00}
\plotone{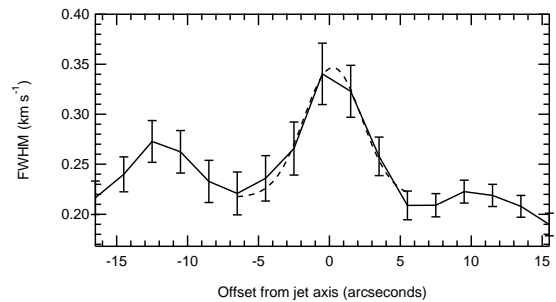}
\caption{NH$_3$ line width (FWHM) as a function of position from the jet axis. The slice used to obtain the profile is shown in Figure~\ref{fig:fwhm}.
Positive (negative) angular units indicate positions northeast (southwest) of the jet axis.  The $1 \sigma$ statistical errors from the spectral fitting are shown for each point.  The profile shows a clear broadening of the line centered on the jet axis.  A Gaussian fit (dashed line) to the central broadened region provides a width (FWHM) of $ 6\arcsec$.  } 
\label{fig:fwhmslice}
\end{figure}

We also see tentative evidence for outflow-envelope interaction in the line width map (Figure~\ref{fig:fwhm}). As discussed in Section~\ref{sec:fwhm}, the region northwest of the source, along the outflow axis, exhibits relatively large line widths that are not correlated with an increase in the gas temperature. This indicates that the line broadening in this region is dominated by non-thermal motions.  The increase in velocity width is spatially correlated with the jet outflow axis (see Figure \ref{fig:fwhm}).  To demonstrate the effect more clearly, we extract the FWHM of the ammonia lines in a slice along the envelope's major axis (i.e., perpendicular to the jet axis) positioned $8 \arcsec$ northwest of the source. The velocity width is averaged over a $4\arcsec$ width, corresponding to the map's synthesized beam.  The result is presented in Figure \ref{fig:fwhmslice}, which shows a statistically significant broadening along the slice.  The observed line width increases from about 0.20 km s$^{-1}$ in the gas not coincident with the jet to approximately 0.35 km s$^{-1}$ at the jet axis. A Gaussian fit to the profile shown in Figure \ref{fig:fwhmslice}, shows that the increase in the velocity width has a FWHM of about $6\arcsec$ ($\sim 1700$ AU), which is barely resolved by our observations, and peaks at the center of the jet axis.

We suggest that the excess in line broadening along the jet axis is  the result of turbulent motions
induced by the interaction of the jet and the dense envelope gas. It is important to note that there is  line broadening throughout the envelope that is not correlated with the jet, and there might be other unknown sources of non-thermal motions causing the broadening.  However, the spatial correlation of the line widths with the jet axis is striking, and suggests that outflow-driven turbulence may be occurring here.  In molecular outflow models where the outflow is driven by a jet formed by a series of bow shocks (as is the case for the HH 211, see Gueth \& Guilloteau 1999), the molecular outflow results from the entrainment of ambient gas in the wake of the bow shocks (Raga \& Cabrit 1993). It is a possibility that the increased velocity dispersion in the ammonia line along the outflow axis is due to the turbulent mixing of entrained gas and the dense gas in the core. It is very probable that the low inclination angle (with respect to the plane of the sky) of the HH 211 outflow  facilitates the detection of this effect in this source. If our interpretation is correct, it would indicate that outflows can drive turbulence inside dense cores, consistent with the results of observational (Swift \& Welch 2008; Arce et al.~2010) and numerical studies (Nakamura \& Li 2007; Carroll et al.~2009) that indicate that outflows can sustain turbulence motions of their surrounding medium. It would also show that outflows may be responsible for the enhanced line width in star-forming ammonia cores (compared to starless cores) observed with single dish telescope surveys (Jijina et al. 1999; Foster et al. 2009).

\subsection{Inner Collapsing Radius} \label{sec:velocity} 

Gas surrounding young stellar objects exhibit different kinematic distributions, depending on the evolutionary status of the central object.  The more evolved systems with less envelope gas often show strong Keplerian signatures, while many younger Class 0 sources appear to reside in elongated and rigidly rotating envelopes.  In either case, kinematics can be complicated by many factors, such as infalling or outflowing gas.  It is also possible that the observed velocities are from optically thick gas that does not represent the motion of the inner layers, or that the mass distribution is not dominated by the central source.

\begin{figure}[h]
\epsscale{1.00}
\plotone{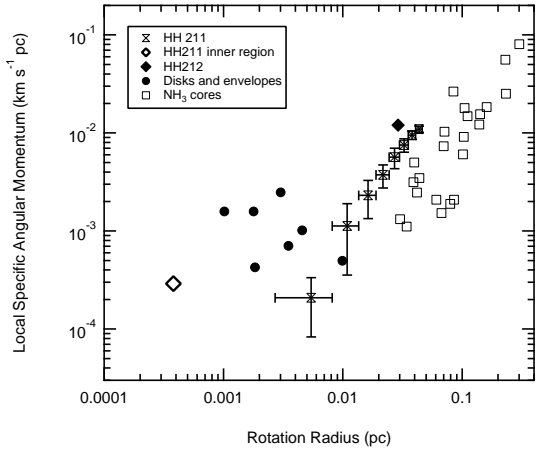}
\caption{Specific angular momentum ($j$) as a function of rotation radius based on Figure 6 of \citet{ohashi1997}.  Points with error bars show values of $j$ in the ammonia envelope of HH 211 for different distances from the protostars.  The error bars show the bin width and the standard deviation of the measurement of $j$ within each bin.For comparison, we also include the point (with an open diamond) corresponding to value reported by Yan et al.~(2010) of the inner rotating structure in HH 211 observed by Lee et al.~(2009). The filled black diamond represents the value of $j$ at the radius of the HH 212 ammonia envelope (from Wiseman et al.~2001). Filled circles include the points for infalling envelope and rotationally supported disks shown in Figure 6 of Ohashi et al.~(1997). Open squares represent the sources from the Goodman et al.~(2003) survey of ammonia cores (also taken from Figure 6 of Ohashi et al.~1997).}
\label{fig:sam}
\end{figure}

The specific angular momentum ($j$) is determined by the internal dynamical motions of the envelope, and by examining the radial profile of $j$, we can infer properties about the state and structure of the envelope.  The specific angular momentum is defined as $j = V_{rot} \times R_{rot}$, where $V_{rot} = V_{obs} / \sin i$ is the rotational velocity at a radius $R_{rot}$, and $i$ is the inclination of the envelope with respect to the line of sight (in our case  $i \sim 85\arcdeg$, see \S~\ref{sec:morphology}). In Figure \ref{fig:sam}, we present the local specific angular momentum in the HH 211 ammonia envelope as a function of distance from the central protostar.  The value of $j$ for each point (along the major axis)  is obtained by averaging over $4\arcsec$-wide bins that  cover the extent of the envelope perpendicular to the major axis at the particular distance from the source.  The error bars show the bin width and the standard deviation of the measurement of $j$ within each bin. Figure \ref{fig:sam} shows that the specific angular momentum of the HH 211 envelope (from about 0.005 to 0.04 pc)  decreases with decreasing distance from the source,  with a power-law dependence of $j$ approximately proportional to   $R_{rot}^{2}$, as expected for solid body rotation.  At a distance of about 0.03 pc, the specific angular momentum of the HH 211 ammonia envelope is consistent with that of the HH 212 ammonia envelope (Wiseman et al.~2001), shown in Figure~\ref{fig:sam} as a filled black diamond. 

In Figure \ref{fig:sam} we also show the diagram of $j$ vs. $R_{rot}$ presented by Ohashi et al. (1997). The square shows the data  from the Goodman et al.~(1993) survey of ammonia cores of different sizes and each point represents a different core, where the value of  $j$  is estimated at the core radius. The cores range in size from about 0.03 pc to 0.3 pc.  A fit to these points shows a power-law dependence of $j$ with radius, with an index of approximately 1.6 (Goodman et al. 1993; Ohashi et al. 1997). This is similar to the dependence of $j$ as a function of distance from HH 211-mm seen in our data. The filled circles in Figure \ref{fig:sam} show both the sources with infalling envelopes and rotationally supported disks in Taurus included in the figure shown by Ohashi et al. (1997). The value of $j$ of these sources is significantly smaller than that of the ammonia cores (open squares) and is approximately constant over a range of sizes (from 0.001 to 0.01 pc). In addition, we also show (with an open diamond) the value of $j$ presented by Yan et al.~(2010) of the  inner rotating structure surrounding HH 211-mm observed by Lee et al.~(2009).

In their study, Ohashi et al. identified two regions with different size dependence of $j$ that diverge at a radius of 0.03 pc: the outer region where cores show a power-law dependence of specific angular momentum with radius; and an inner region where $j$ is relatively constant for small envelopes and disks. They argue that the radius where the two zones divide---the radius inside which the specific angular momentum is conserved---represents the size scale for collapse. Presumably cores in the power-law are mainly in virial equilibrium, and the bulk of the gas is not currently undergoing dynamical collapse.   
The points representing the value of $j$ in the ammonia envelope of  HH 211 are basically consistent with the trend seen in the ammonia cores observed by Goodman et al. (1993). That is, it appears that the gas probed by our ammonia observations (at a distance from HH 211-mm from about 0.005 
to 0.04 pc) is gravitationally stable. Indeed, a comparison of the turbulent, rotational and gravitational energies of the entire ammonia envelope mapped by our VLA observations show that the gas is close to virial equilibrium. 

Comparing our data with results from previous studies, it appears that the gas within approximately 0.005 pc of the central protostar is undergoing dynamical collapse. The specific angular momentum associated with the inner rotating structure observed by Lee et al.~(2009) is larger than would be expected if we were to extrapolate the relation we observe in the ammonia envelope. This is similar to the relation between the $j$ values for the ammonia cores and the infalling envelope and disks in the Ohashi et al. plot. It therefore appears that the inner rotating structure does not share the dynamical properties of the ammonia envelope, and most likely is inside the region where dynamical collapse  is taking place. Inside the radius for dynamical collapse in a core, the specific angular momentum should be constant,  which implies that the value of $j$ of the inner rotating structure in HH 211 should be similar to that of the outermost radius of the region of dynamical collapse. From Figure \ref{fig:sam} we see that a value of $j$ in the ammonia envelope similar to that of the inner rotating structure in HH 211 is reached at a distance from HH 211-mm of about 0.005 pc. Following Ohashi et al., this suggests that the size scale for collapse in HH 211 is about 0.005 pc (a factor of 6 smaller than what they found in the Taurus sources) and the gas within this radius is undergoing dynamical collapse, while the gas at larger distances from the source is gravitationally stable.

The size and evolution of the inner collapsing radius will have a profound effect on the final mass of the central object. Using our ammonia observations, we estimate that there is only about $0.02 \mbox{ M}_\odot$ of gaseous material within 0.005 pc of HH 211-mm. This implies that if the collapsing radius maintains its current size until the end of the mass infall process in the HH 211 envelope, then the central protostar(s) will accumulate no more than  $\sim 0.02 \epsilon$ M$_\odot$, where $\epsilon$ is the star formation efficiency factor or, in other words, the fraction of infalling material that ends up in the forming star(s). In the HH 211-mm system, the primary binary companion is estimated to  currently have a mass of about 0.04 M$_\odot$. Adopting a star formation efficiency parameter of $\epsilon \sim 0.3$, and assuming that the collapsing radius does not grow, and that the ratio of matter that accretes into each companion is the same as their mass ratio, then the final mass of the primary companion will be no more $\sim 0.05 \mbox{ M}_\odot$ (a brown dwarf). Therefore if the primary companion is to eventually become a stellar object (of M $> 0.08 \mbox{ M}_\odot$) at the end of the mass-assembling stage, then the collapsing radius would have to grow by about a factor of six (i.e., to $\sim 0.03$ pc) to encompass a volume large enough to have the necessary amount of material that can infall to produce a hydrogen-burning star in the center. It is possible that the collapsing radius increases as the embedded protostar evolves (see Yan et al.~2010 for a discussion on this), and since HH 211-mm is believed to be a young (Class 0) protostar there is still time for the collapsing radius to grow. 
Knowing what sets the collapsing radius of an envelope and how it evolves is clearly essential for understanding how stars acquire their mass.

\section{Summary and Conclusion}

We have studied the gas of the envelope surrounding HH 211-mm using simultaneous observations of the $\mbox{NH}_3$ (1,1) and (2,2) line emission.   The integrated intensity and column density maps reveal an elongated structure of about $10^4$~AU that is perpendicular to the outflow.  Within the envelope, we find three peaks in the column density, one of which is probably associated with the dense inner envelope surrounding the central source,  and the other peaks might be the result of fragmentation.   The velocity distribution of the envelope clearly exhibits a  gradient of approximately $6.3$ km s$^{-1}$ pc$^{-1}$ along the axis perpendicular to the jet.   We find that the HH 211 envelope contains a large amount of rotational energy, as  the ratio of the rotational kinetic energy to the gravitational energy is relatively high ($\beta_{rot} \sim 0.2$), and we suggest this may be the cause of the fragmentation that is observed in the column density map.  

The ammonia line broadening in the envelope is the result of both thermal and non-thermal contributions.  Because we have measured the kinetic temperature, we are able to identify regions where non-thermal broadening is dominant. Our results clearly show that the envelope is mostly internally heated and has a peak kinetic temperature near the central source. Close to the protostar and other regions where the kinetic temperature is high we find  that the thermal width is comparable to the non-thermal width, which indicates that the line broadening can be partially attributed to increased gas temperature caused by radiation from the central protostar. However, additional sources of non-thermal motions are needed to explain the increased line broadening at the center of the envelope, and  we propose these are  caused by a combination of unresolved outflow and infall motions. 

We find evidence for outflow impact on the dense envelope gas. There is a clear gradient along the minor axis of the envelope, which is coincident with the HH 211 jet axis. The gradient is in the same sense as that of the outflow, with blueshifted velocities to the southeast and redshifted velocities to the northwest of the source. We estimate that the kinetic energy injected by the molecular outflow is sufficient to produce the velocity gradient in the high-density gas, and if the current trend continues, the HH 211 outflow could eventually clear the dense gas along the jet axis. Evidence for outflow-envelope interactions is also seen in the large line width to the northwest of the source, along the outflow axis. 
In contrast to the central envelope regions, the increased line widths along the outflow axis are not accompanied by high kinetic temperature and thus must come from an increase in the non-turbulent motions. We suggest the increase in the line width is produced by  turbulence induced by the jet interaction with the envelope gas.  

From our study of the HH 211 ammonia envelope and data from the literature we argue that the 
 envelope has an inner collapsing region (within 0.005 pc from  the source) while the rest of the envelope is  dynamically stable.  Evidence for this scenario is found in the profile of the  specific angular momentum as a function of the separation from the central protostar.  In the ammonia envelope (from about 0.005 to 0.4 pc), we find the angular momentum to be clearly increasing with radius. However, the specific angular momentum of the inner rotating structure (at about 80 AU) reported by Lee et al.~(2009) has a value close to the angular momentum  of the innermost radius we can probe with our VLA ammonia observations (0.005 pc). Following the study by Ohashi et al.~(1997), we suggest that the specific angular momentum must be nearly constant at approximately $3 \times 10^{-4}$ km$^{-1}$ pc$^{-1}$ within 0.005 pc. This inner region, where angular momentum is conserved, represents the (current) size of the collapsing region surrounding the central protostar(s). If we assume that the central source (or sources) have a total mass of about $0.05$ M$_\odot$ (as indicated by Lee et al. 2009), we find that  there is not enough mass within the current collapsing radius of $0.005$ pc to eventually produce a stellar object in the HH 211 system. The collapsing radius will have to grow by about a factor of six for there to be enough mass  to eventually produce a star.  It is clear that how the inner collapsing region evolves will have an impact on the final outcome of the object forming in the center. A study of the evolution of the collapsing region in a sample of star-forming cores will be essential for understanding  the mass-assembling process in protostars.

\acknowledgments
The authors wish to thank the referee for insightful suggestions that improved the manuscript. We  thank Noami Hirano for providing us with the H$_2$ image of the HH 211 jet. JT is partially supported by a PGS-D scholarship of the Natural Sciences and Engineering Research Council of Canada. This is study was supported by NSF award AST-0845619 granted to HGA. 

{\it Facilities:} \facility{VLA}.

\clearpage

\end{document}